\documentclass[]{jpsj3}
\usepackage{graphicx}% Include figure files
\usepackage{dcolumn}% Align table columns on decimal point
\usepackage[usenames]{color}
\usepackage{bm}% bold math

\title{Low-temperature magnetic properties of the Kondo lattice model in one dimension}

\author{Sorato Minami and Hikaru Kawamura\thanks{E-mail:kawamura@ess.sci.osaka-u.ac.jp}}

\inst{
Department of Earth and Space Science, Graduate School of Science, Osaka University, Toyonaka 560-0043, Japan
}%

\abst{
Low-temperature magnetic properties of the one-dimensional Kondo-lattice model with classical localized spins are investigated by means of a Monte Carlo simulation combined with an exact diagonalization technique. Comparative simulations are also made for the RKKY classical Heisenberg model known to be perturbatively derived in the weak-coupling limit. In addition to the previously identified antiferromagnetic, ferromagnetic and coplanar helical phases, the chiral noncoplanar phase and the collinear phases of the period-4 and 6 are newly identified, neither of which is realized in the corresponding RKKY model. The period-4 collinear phase persists even in the weak-coupling limit at the 1/4- and 3/4-fillings, invalidating the perturbative approach at these fillings. 
}

\kword{Kondo lattice model, RKKY model, one dimension, noncoplanar state, helical state, scalar chirality, Monte Carlo simulation}

\begin{document}
\maketitle

\section {Introduction}

 Itinerant electrons interacting with localized moments has provided a fruitful stage for a variety of intriguing phenomena in condensed matter physics. The Kondo-lattice model (KLM), sometimes referred to as the $s-d$ model or the Hund model, is the simplest model to describe such an interplay between conduction electrons and localized spins. The model has intensively been studied in the past in the context of, {\it e.g.\/}, heavy fermions and manganites \cite{Tsunetsugu,Hamada,Yunoki,Yunoki2,Dagotto,Motome00,Chattopadhyay,McCulloch,Garcia,Basylko,Pradhan,Peters,Peters13,Misawa}, and more recently, in geometrically frustrated systems \cite{Martin,Akagi,Kato,Akagi12,Motome10,Chern,Ishizuka,Ghosh,Barros} and in the context of Dirac electrons \cite{Barros,Hayami}.

 The spin quantum number $S$ of the localized spin is a parameter of the model depending on the target material. It is $S=3/2$ for manganite and could be as large as $S=7/2$ for Gd. In such cases, the classical approximation for the localized spin might be reasonable. By contrast, the $S=1/2$ localized spin describes a highly quantum mechanical case where quantum spin fluctuations and the Kondo-singlet formation might play an important role \cite{Tsunetsugu,McCulloch,Garcia,Basylko,Motome10,Peters,Peters13,Misawa}. In the present paper, we concentrate on the case of the classical localized spin. Even in such a classical KLM, the determination of its magnetic, thermodynamic and transport properties and the determination of the phase diagram has been a nontrivial task.

 Possible thermodynamic phases and the phase diagram of the classical KLM on the cubic lattice has been studied by various authors \cite{Hamada,Yunoki,Yunoki2,Dagotto,Motome00,Chattopadhyay,Pradhan,Hayami}. Concerning the $T=0$ phase diagram, the model exhibits an antiferromagnetic (AF) order at zero chemical potential $\mu=0$ or at half-filling $n=1$ ($n$ the conduction-electron number density summed over up and down spins), and exhibits a ferromagnetic (F) or an incommensurate helical order away from the half-filling. The transition between the AF phase and the F phase is of first-order, {\it i.e.\/}, a phase separation occurs between the uniform states.

 In the weak coupling limit where the exchange coupling $J$ between the conduction-electron spin and the localized spin is much smaller than the transfer energy  $t$ of the conduction electron, the second-order perturbation analysis reduces the original KLM to the localized-spin-only model, the so-called RKKY model, where the Heisenberg spins interact via the long-range and oscillating RKKY interaction. In the opposite limit of the strong coupling $J>>t$, the model is reduced to the so-called double-exchange model. In this regime, the ground state of the model tends to be ferromagnetic. In the intermediate parameter range of $J\sim t$ where the nontrivial interplay between the conduction electrons and the localized spins are expected, a theoretical treatment becomes more difficult. In this regime, one usually needs to resort to some numerical method to treat both the conduction-electron and the localized-spin degrees of freedom simultaneously. 

 Recently, renewal of interest in the KLM has occurred via the studies of geometrically frustrated lattices such as the triangular \cite{Martin,Akagi,Kato,Akagi12}, the kagome \cite{Ghosh,Barros} and the pyrochlore lattices \cite{Chern,Ishizuka}. In particular, the KLM was reported to yield exotic spin states, which are not realized in the corresponding RKKY spin-only model. For example, the KLM on the pyrochlore lattice, though exhibiting the ``all-in, all-out''- or ``2-in, 2-out''-type spin structures similarly to those exhibited by the corresponding RKKY spin-only model \cite{Ikeda}, exhibits at certain fillings exotic states never exhibited by the RKKY model \cite{Chern,Ishizuka}.  A rich variety of noncoplanar chiral states were also reported in the KLM on the triangular lattice at the 1/4- and the 3/4-fillings \cite{Martin,Akagi,Kato,Akagi12}, or in the one on the kagome lattice \cite{Ghosh,Barros}.

 Under such circumstances, it is most interesting to further clarify the situation where the nontrivial noncoplanar spin order arises in the KLM, as with the relation between the KLM and the RKKY models. It would be natural to expect that a moderately large $J$  plays an essential role in stabilizing the noncoplanar spin structure. Yet, the possible dependence on the conduction-electron density $n$ (or the chemical potential $\mu$) remains highly nontrivial. Furthermore, even in the weak coupling limit $J/t\rightarrow 0$, the validity of the RKKY model is not completely obvious, if some non-perturbative effects would become relevant.
% In such a case, the perturbative scheme itself within which the RKKY model has been derive may fail. 

 In order to examine these issues, we wish to investigate in the present paper the properties of the one-dimensional (1D) KLM with classical Heisenberg spins, in comparison with the corresponding properties of the 1D RKKY classical Heisenberg model. Indeed, the 1D KLM with the classical spin might be the simplest model in this category amenable to a full numerical analysis, where we can map out its detailed phase diagram and make a close comparison with the corresponding properties of the RKKY spin model. 
%Our motivation is twofold. First, the 1D KLM with the classical spin is the simplest one amenable to a full numerical analysis where we can map out its detailed phase diagram and make a close comparison with the corresponding properties of the RKKY spin model. Second, the model might have some relevance to the recent experiment on ** where the quasi-1D conduction electrons at * are coupled to localized moments at *.

 We perform Monte Carlo (MC) simulations on the localized classical spin degrees of freedom, tracing out the conduction-electron degrees of freedom by the exact diagonalization (ED) method. We then investigate its low-temperature magnetic structures of localized spins and determine the $T\rightarrow 0$ phase diagram as a function of both the coupling $J$ and the chemical potential $\mu$. We succeed in elaborating the phase diagram of the model. We also perform MC simulations on the corresponding 1D RKKY Heisenberg model. The low-temperature spin structures of the two models are then compared and discussed.

 It is found that the 1D RKKY model always exhibits the coplanar helical spin order in the $T\rightarrow 0$ limit. Such helical orders are generally incommensurate with the underlying lattice. Although the KLM is expected to reduce to the RKKY model in the weak-coupling limit, and we indeed observe such coplanar helical states consistent with the ones stabilized in the RKKY model, we observe around the 1/4- and 3/4-fillings the period-4 collinear state, an ``up-up-down-down'' state, never realized in the RKKY model. Just at the 1/4- and 3/4-fillings, this collinear state remains stable even in the $J/t\rightarrow 0$ limit, revealing that the perturbation approach fails at these fillings. We identify the cause of this failure as a spontaneous gap opening occurring there. In the intermediate-coupling regime, nontrivial chiral noncoplanar spin structures, again never realizable in the corresponding RKKY model, are stabilized in a wide range of the phase diagram. These noncoplanar spin configurations often look complicated in real space, but the underlying hidden periodicity becomes eminent when one looks at the spin scalar chirality. Further away from the half-filling, the F state is stabilized.

 In section 2, we introduce our models, the 1D KLM and the 1D RKKY Heisenberg model. The computational method employed is also explained. The results of our numerical simulations on the 1D RKKY Heisenberg model are presented in section 3. In section 4, the results of our numerical simulations on the 1D KLM are presented, with particular interest in its low-temperature spin structure and the phase diagram. Comparison is made with the results of the corresponding RKKY Heisenberg model. Section 5 is devoted to summary and discussion.

\section {The model and the method}

The model of our primary concern is the 1D KLM with the classical Heisenberg spin, whose Hamiltonian is given by
\begin{equation}
{\cal H} = -t \sum_{<ij> \sigma} (c^\dag_{i\sigma} c_{j\sigma} + h.c.) - J \sum_i \vec S_i\cdot \vec s_i -\mu \sum_{i\sigma} c^\dag_{i\sigma} c_{i\sigma} ,
\end{equation}
where $c^\dag_{i\sigma}$ and $c_{i\sigma}$ are the creation and the annihilation operators of a conduction electron at the site $i$ and of the spin $\sigma$, $\vec s_i=\frac{1}{2} c^\dag_{i\sigma} \vec \tau_{\sigma\sigma'} c_{i\sigma'}$ ($\vec \tau$ the Pauli matrices) is the conduction-electron spin, $\vec S_i$ is the three-component classical Heisenberg spin at the site $\vec S_i=(S_{ix}, S_{iy}, S_{iz})$ with $|\vec S_i|=1$, $t>0$ the electron transfer, $J$ the exchange coupling, $\mu$ the electron chemical potential, and the sum $<ij>$ is taken over all nearest-neighbor pairs on the 1D lattice. For the classical spin, the sign of the exchange $J$ is actually irrelevant. We assume $J$ to be ferromagnetic $J>0$ here.

The partition function $Z$ is given by 
\begin{equation}
Z = {\rm Tr}_{c^\dag_{i\sigma}, c_{i\sigma}} {\rm Tr}_{\vec S_i} e^{-\beta {\cal H}} ,
\end{equation}
where $\beta=1/T$ is the inverse temperature and the trace is taken over both the conduction-electron and the localized-spin degrees of freedom. By taking the partial trace only over the electron degrees of freedom for a given spin configuration \{$\vec S_i$\}, one gets an effective Hamiltonian for the spin degrees of freedom ${\cal H}_{eff}(\{\vec S_i \}; \beta)$, defined by
\begin{equation}
e^{-\beta {\cal H}_{eff}(\{\vec S_i\}; \beta)} = {\rm Tr}_{c^\dag_{i\sigma}, c_{i\sigma}} e^{-\beta {\cal H}} .
\end{equation}
Note that the effective spin Hamiltonian ${\cal H}_{eff}$ defined in this way is temperature dependent.

 The second-order perturbation in terms of $J/t$, with an additional assumption of the spherical Fermi surface of conduction electrons, leads to an effective spin-only Hamiltonian,
\begin{equation}
H_{RKKY} = - \sum_{<ij>} J_{RKKY}(r_{ij}) \vec S_i\cdot \vec S_j ,
\end{equation}
where $r_{ij}=|i-j|$ is the lattice distance between the sites $i$ and $j$. In 1D, this RKKY coupling $J_{RKKY}$ is given by \cite{Tamura,Yafet,Litvinov}
\begin{equation}
J_{RKKY}(r) = J_0 \int^\infty _r \frac{\sin (2k_Fx)}{x} {\rm d}x,
\end{equation}
where $J_0$ is the energy constant. The RKKY Hamiltonian is characterized only by one parameter $k_F$, the Fermi wavenumber.

 For the RKKY spin-only model, we employ the standard MC technique, the Metropolis method. The total number of spins $L$ is $L=100$, periodic boundary condition being applied. In determining the low-temperature spin configuration, we examine the finite-size effect by simulating larger sizes of $L=200,300,500$, and the effect of the boundary condition by employing the free boundary condition as well. Note that, while we have not performed the Ewald summation, the RKKY interaction is summed over all spin-pairs on the lattice without any cut-off introduced. Relatively large lattice sizes of $L\leq 500$ employed in our simulations and a systematic analysis of the finite-size effect on the result would guarantee that the truncation error associated with the long-range RKKY interaction gives a negligible effect on our results.

 Physical quantities are computed as a function of the temperature $T$ by gradually cooling the system from the high temperature. Total number of MC steps is $4\times 10^5$ sweeps at each temperature, the first $2\times 10^5$ discarded for thermalization and the subsequent $2\times 10^5$ used in computing physical quantities. Below, the energy and the temperature are measured in units of $J_0$. The spin configuration in the low-temperature limit is taken at a low temperature of $T=10^{-5}$ to eliminate the thermal noise.

 For the KLM, the trace over classical Heisenberg spins is taken by the MC method, the Metropolis method, under the temperature-dependent effective spin Hamiltonian ${\cal H}_{eff}$. The electron trace in evaluating ${\cal H}_{eff}$ is taken by the ED technique. The total number of spins is $L=20\sim 100$, periodic boundary condition being applied.  In determining the low-temperature spin configuration, we also examine the effect of the boundary condition by employing the free boundary condition as well. Physical quantities are computed as a function of the temperature $T$ by gradually cooling the system from the high temperature, with the coupling $J$ and the chemical potential $\mu$ as the parameters. Total number of MC steps is $4\times 10^5$ sweeps at each temperature, the first $2\times 10^5$ discarded for thermalization and the subsequent $2\times 10^5$ used in computing physical quantities. Below, the energy, the coupling $J$, the chemical potential $\mu$ and the temperature are all measured in units of $t$. The spin configuration in the low-temperature limit is taken at a low temperature of $T=10^{-5}$ to eliminate the thermal noise.
%Physical quantities are computed as a function of the temperature $T$, the chemical potential $\mu$, and the coupling ratio $J/t$, paying attention to the fact that the effective ${\cal H}_{eff}$ employed in MC estimates is actually temperature-dependent. 

 As mentioned, the transition of the KLM is often of first-order, accompanied by a phase separation. In the present paper, we are mainly concerned with the nature of the uniform phases of the model, and compute physical quantities and construct a phase diagram as a function of the chemical potential $\mu$, rather than of the electron number density $n$. From a simulation standpoint, this serves to avoid possible complications associated with the phase separation and severer finite-size effects associated with the domain or the interface. Of course, either the chemical-potential ($\mu$) given or the electron-density ($n$) given is just the choice of the ensemble, and does not affect any physical property of the model including its phase structure. The phase-separate (two-phase coexisting) state, which appears in the $n-T$ phase diagram occupying a finite fraction of it, is shrunk onto the first-order transition line in the corresponding $\mu-T$ phase diagram. Yet, the phase structure of the model is common between the two ensembles: : all the uniform phases realized in the ($\mu,T$) ensemble appears also in the ($n,T$) ensemble, in which the phase-separate state consists of the two uniform states realized in the ($\mu, T$) ensemble.

 Because of the electron-hole symmetry of the model,  the phase diagram should be symmetric with respect to $\mu \leftrightarrow -\mu$. Hence, we can restrict $\mu$ to either sign without loosing generality. In the following, we assume $\mu$ to be non-negative $\mu\geq 0$, and also $n\geq 1$ ($n$ in the $\mu<0$ region is obtained by $n=2-n$).

\section{The results on the RKKY Heisenberg model}

\begin{figure}
 \includegraphics[width=8cm]{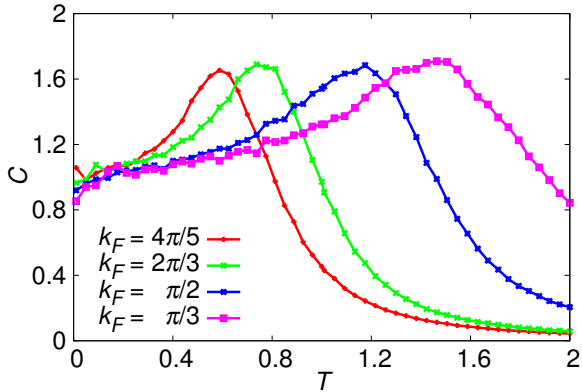}
 \caption{(color online). The specific heat per site of the one-dimensional RKKY Heisenberg model plotted versus the temperature for several values of the Fermi wavenumber $k_F$. The lattice size is $L=100$.
}
\end{figure}

 In this section, we report on our MC results on the 1D RKKY Heisenberg model. Fig.1 exhibits the temperature dependence of the specific heat per site for the size $L=100$ for several values of the Fermi wavenumber $k_F$. The data exhibit a broad single peak associated with the onset of the spin short-range order. 

 Some of the stable low-temperature spin configurations determined by the gradual cooling procedure are illustrated in Fig.2 for a commensurate and for an incommensurate $k_F$-values, {\it i.e.\/}, (a) $k_F=\pi/2$, and (b) $k_F=2.55$. The stable low-temperature spin configurations turn out to be always coplanar spirals of the wavenumber $k=k_F$. For general $k_F$, the spiral is incommensurate with the underlying lattice, a typical example shown in Fig.2(b). By contrast, for a commensurate $k_F$, a commensurate spiral is realized, a typical example shown in Fig.2(a) where the spiral is of the period-4, {\it i.e.\/}, the 90-degrees spiral.

\begin{figure}
 \includegraphics[width=8cm]{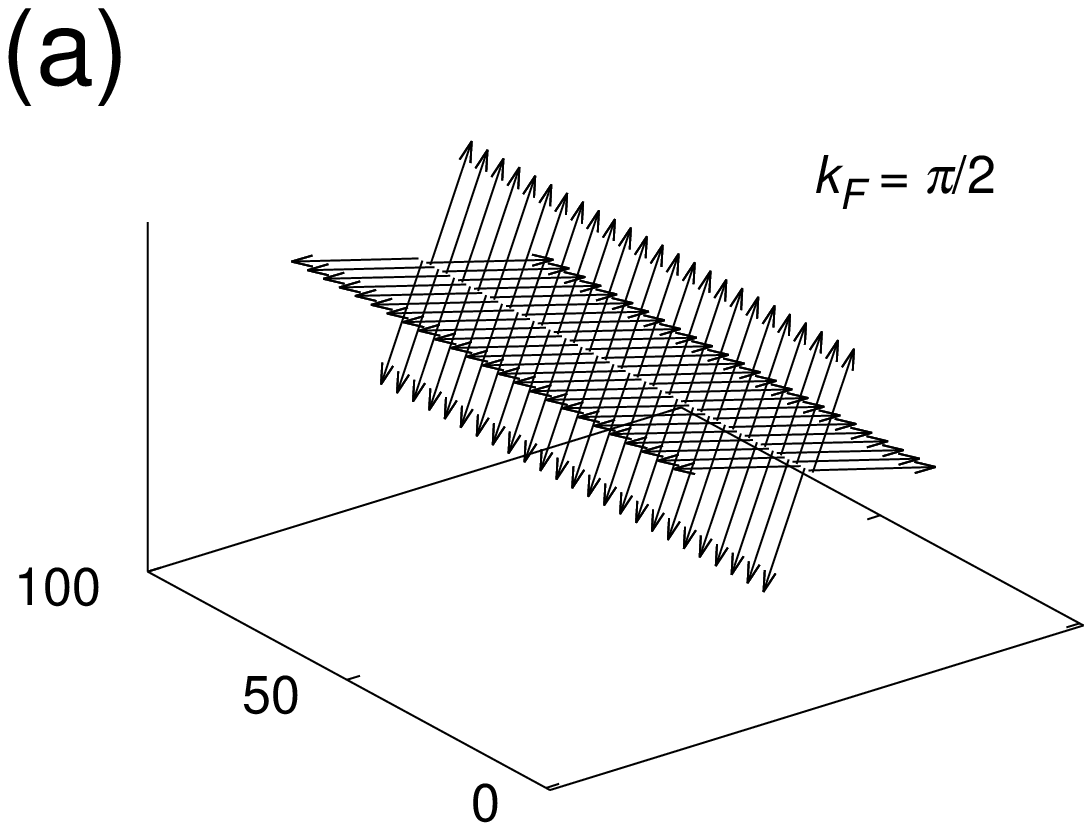}
 \includegraphics[width=8cm]{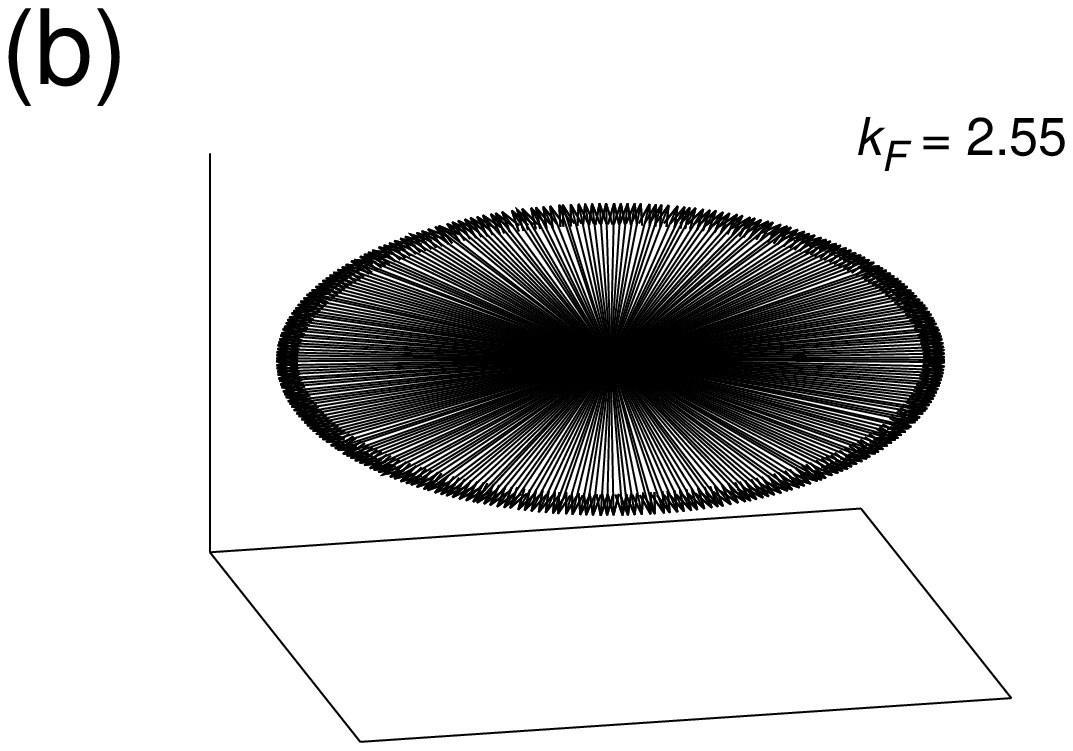}
 \caption{The low-temperature spin configurations of the one-dimensional RKKY Heisenberg model for (a) $k_F=\pi/2$, and (b) $k_F=2.55$, obtained by gradually cooling the system to $T=10^{-5}$. The lattice size is $L=100$ for (a), and $L=300$ for (b). In (b), the origin of all the spin vectors is collected to a common point for illustration.
}
\end{figure}

We note that, when the system size is small, noncoplanar spin structures sometimes show up as a finite-size effect, since the combination of the small lattice size and the applied periodic boundary condition sometimes forces the system to take such a configuration. Even in such a case, large enough sizes always lead to the coplanar spirals. We also confirm that the free boundary condition always induces the coplanar spin configuration in its interior. Hence, the stable spin structure of the 1D RKKY model turns out to be rather simple, {\it i.e.\/}, the coplanar spiral of the wavevector $k_F$.

\section{The results on the Kondo lattice model}

In this section, we report on our numerical results on the 1D KLM. In Fig.3, we show the temperature dependence of the specific heat per site in the case of $J=1$ for several values of the chemical potential $\mu$.  The data exhibit two well-separated peaks. The upper one is  associated with the conduction-electron degrees of freedom, while the lower one with the onset of the localized-spin short-range order. 

\begin{figure}
 \includegraphics[width=8cm]{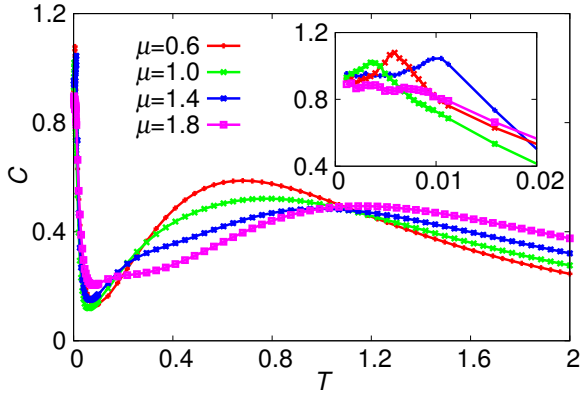}
 \caption{(color online). The specific heat per site of the one-dimensional Kondo-lattice model with $J=1$ plotted versus the temperature for several values of the chemical potential $\mu$. The system size is $L=20$. The inset is a magnified view of the lower-temperature peak region.
}
\end{figure}

\subsection{Phase diagram}

We examine the stable localized-spin configurations in the $T\rightarrow 0$ limit by the numerical procedure explained in \S II for various $J$ and $\mu$, and determine the $T\rightarrow 0$ phase diagram in the $J$-$\mu$ plane. The result is shown in Fig.4. We restrict the region to $\mu\geq 0$ ($1\leq n\leq 2$) making use of the electron-hole symmetry. The phase diagram is completely symmetric with respect to $\mu \leftrightarrow -\mu$ ($n \leftrightarrow 2-n$).

 We note that we take the chemical potential $\mu$, not the electron number density $n$, as an independent parameter. Because of this, the phase diagram might have an appearance somewhat different from the one where $n$ is taken as an independent parameter. For example, although the AF state, {\it i.e.\/}, an ``up-down'' state, takes a large portion of the phase diagram of Fig.4, the electron number density is always $n=1$ there, {\it i.e.\/}, strictly half-filled, so that the AF state is shrunk to a single line in the $n-J$ phase diagram (to the $n=1$ line). Phase-separation regions are not visible in Fig.4, only uniform phases appearing.

\begin{figure}
 \includegraphics[width=9.5cm]{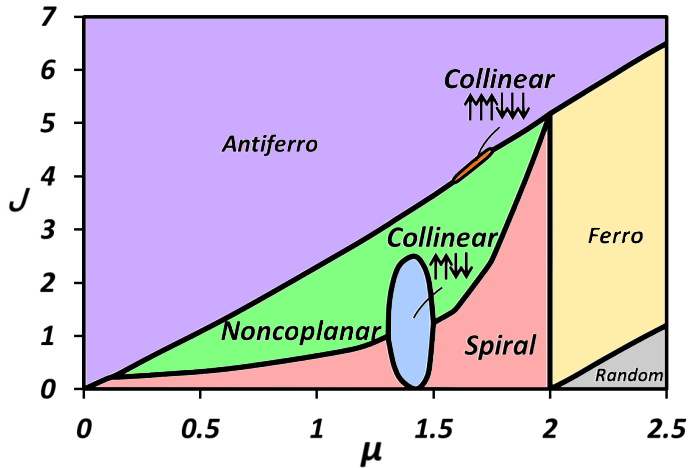}
 \caption{(color online). The $T\rightarrow 0$ phase diagram of the one-dimensional Kondo-lattice model in the chemical potential $\mu$ versus the coupling $J$ plane. Both $\mu$ and $J$ are normalized by the electron transfer $t$.
}
\end{figure}

 In addition to the AF, F and incommensurate helical phases previously reported, some other phases are identified in Fig.4, such as the collinear phases of the period-4 and 6, and the chiral noncoplanar phase.

 In the weak-coupling regime $J<<1$, we generally observe coplanar spiral spin structures, essentially of the type observed in the RKKY model. Noncoplanar spin structures sometime arise for small systems, but they give way to the coplanar spirals for larger systems as in the case of the RKKY model. For general values of $\mu$, the coplanar spiral is incommensurate with the underlying lattice. An example is shown in Fig.5(a).

 For special values of $\mu$ close to $\mu=\sqrt 2$ corresponding to the 3/4-filling ($n=3/2$), on the other hand, the collinear ``up-up-down-down'' structure of the period-4 turns out to be stabilized even in the weak-coupling regime. An example is shown in Fig.6(a). Due to the electron-hole symmetry, exactly the same thing occurs at the 1/4-filling ($n=1/2$). Just at $\mu=\sqrt 2$, this collinear phase remains stable even in the limit $J\rightarrow 0$, indicating that the RKKY model derived from the perturbation approach fails to capture the correct physics at these particular fillings. We shall further discuss the reason of this failure of the perturbation approach below.

 For all other values of $\mu\leq 2$, we find that the coplanar spiral as expected in the RKKY model is stabilized for sufficiently small $J$, indicating that the perturbation approach is justifiable for  sufficiently small $J$. It should be noticed, however, that the collinear spin structure of the period-6, ``up-up-up-down-down-down'' structure as demonstrated in Fig.7(c), is stabilized in the intermediate-coupling regime of $J\simeq 4$ near $\mu=\sqrt 3$ in a narrow window of the phase diagram, though the coplanar spiral of the period-6, the 60-degrees spiral, is stabilized for smaller $J$. Other collinear structures, {\it i.e.\/}, the one with longer periods or with odd-number periods, is never observed as a stable state in our simulations.

 For $\mu$ greater than two, there also appears a ``disordered'' spin state in the weaker-coupling regime. In this state, the site is fully occupied by conduction electrons, {\it i.e.\/}, $n=2$. No hopping is allowed for the conduction electrons so that the localized spins are effectively free in this state.

 For intermediate $J$, we generally observe chiral noncoplanar structures as demonstrated in Figs.5(b), 6(b) and 7(b). We carefully check the size dependence of the structure, to find that the noncoplanar structure observed in the intermediate-coupling regime is not the finite-size effect. In fact, it is often the case that the coplanar spiral observed in smaller systems gives way to the noncoplanar structure as the system size is increased. Further confirmation is obtained from the simulation under the free boundary condition, where similar noncoplanar spin structure is realized even in the deep interior of the system. The observed structures are generally incommensurate with the underlying lattice, and apparently look quite complex in real space, as can be seen from Figs.5(b), 6(b) and 7(b). We shall further discuss below the hidden periodicity of these complex structures.

\begin{figure}
 \includegraphics[width=8cm]{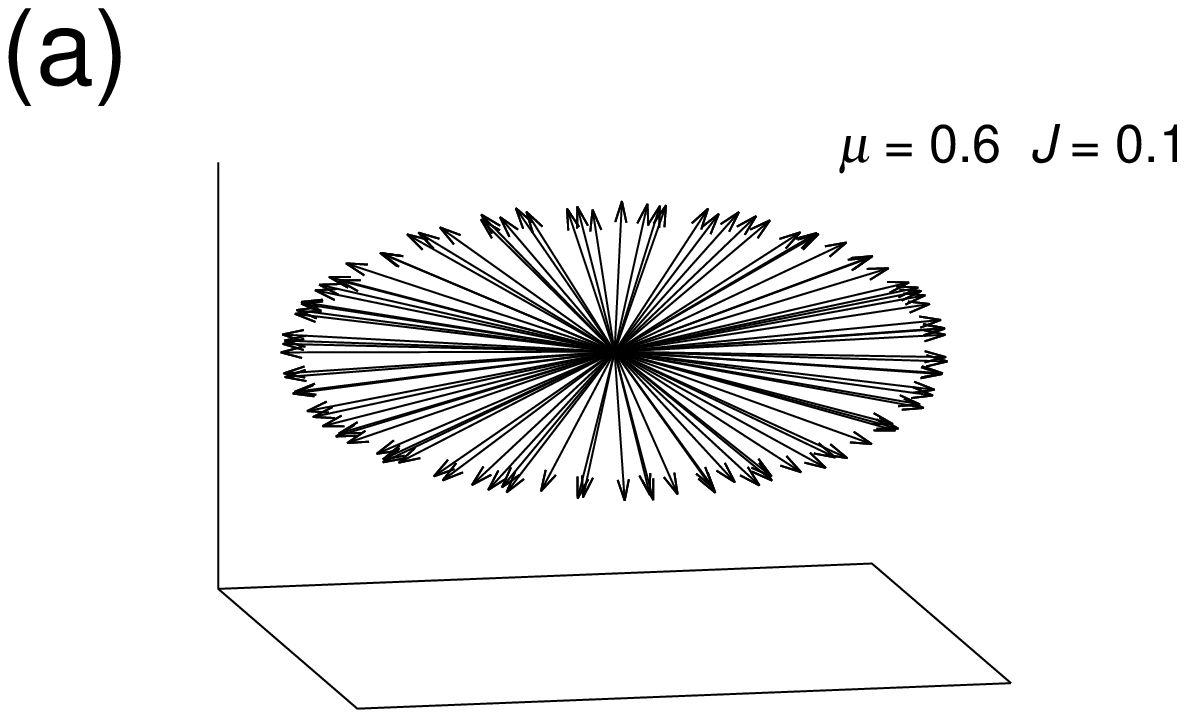}
 \includegraphics[width=8cm]{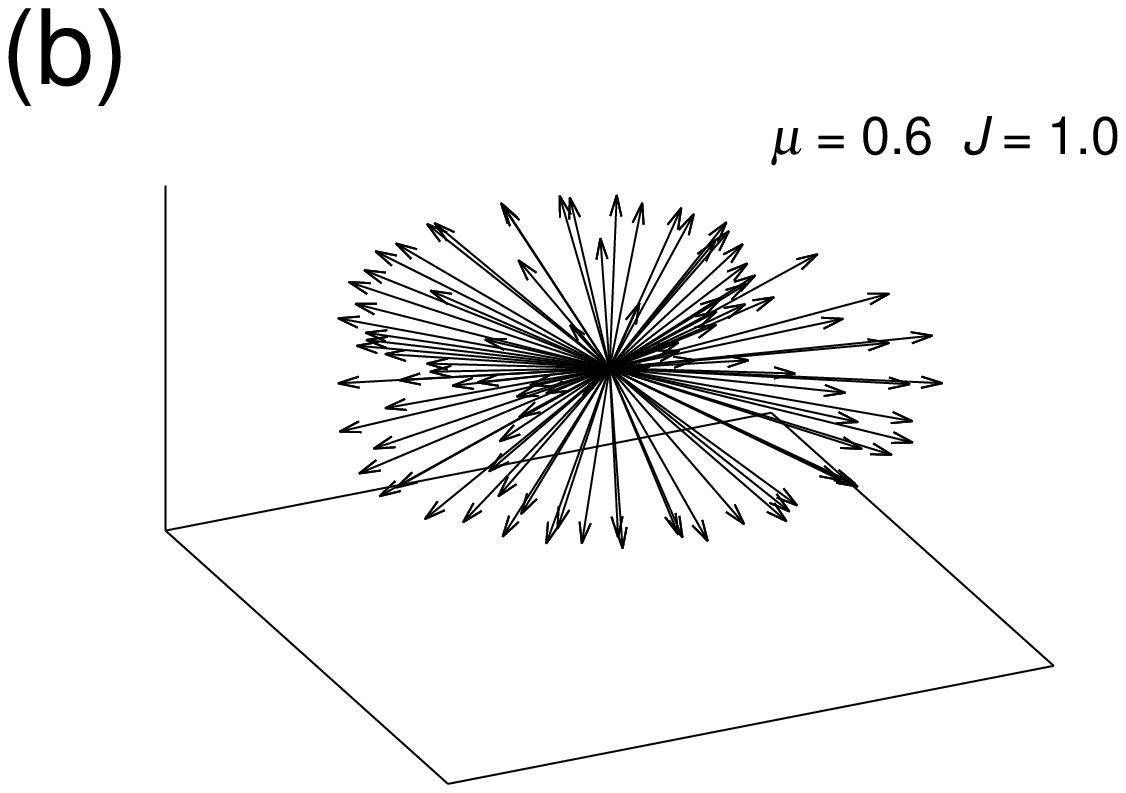}
 \caption{The low-temperature localized-spin configurations of the one-dimensional Kondo-lattice model at $\mu=0.6$, obtained by gradually cooling the system to $T=10^{-5}$. The coupling is (a) $J=0.1$, and (b) $J=1.0$, each lying in the coplanar helical and the noncoplanar chiral regimes, respectively. The lattice size is $L=100$. The origin of all the spin vectors is collected to a common point for illustration.
}
\end{figure}

 For sufficiently large $\mu$ or large $n$, the F state is stabilized. The transition into the F state from the noncoplanar or the AF state is of first-order, accompanied by a discontinuous jump in the magnetization and in the electron number density $n$ ($1.4\lesssim n<2$ in the F phase). The AF-F first-order transition in the strong coupling regime at $J\gtrsim 5$ is the one intensively discussed in the literature.\cite{Yunoki,Yunoki2,Dagotto,Motome00,Chattopadhyay,Pradhan} In contrast, in the intermediate- or in the weak-coupling regime, the transition is more complex and of multi-steps. The chiral noncoplanar state, the coplanar spiral state, and sometimes even the collinear state appear in between.

\begin{figure}
 \includegraphics[width=8cm]{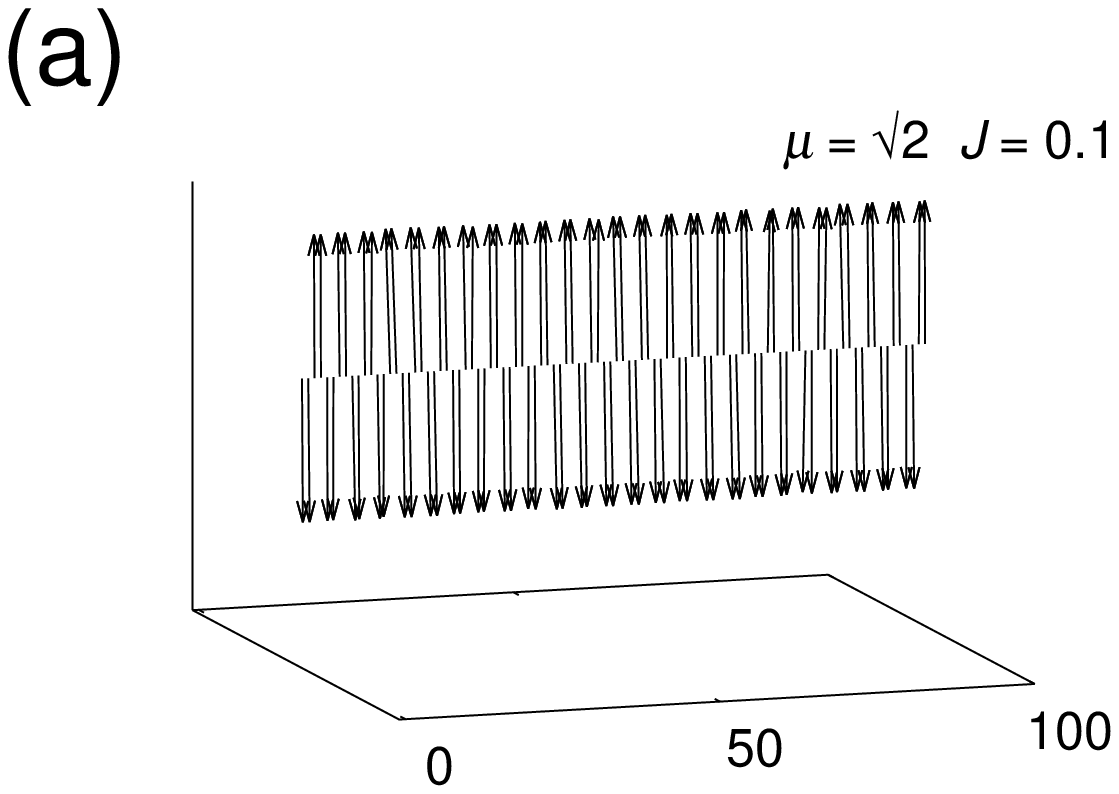}
 \includegraphics[width=8cm]{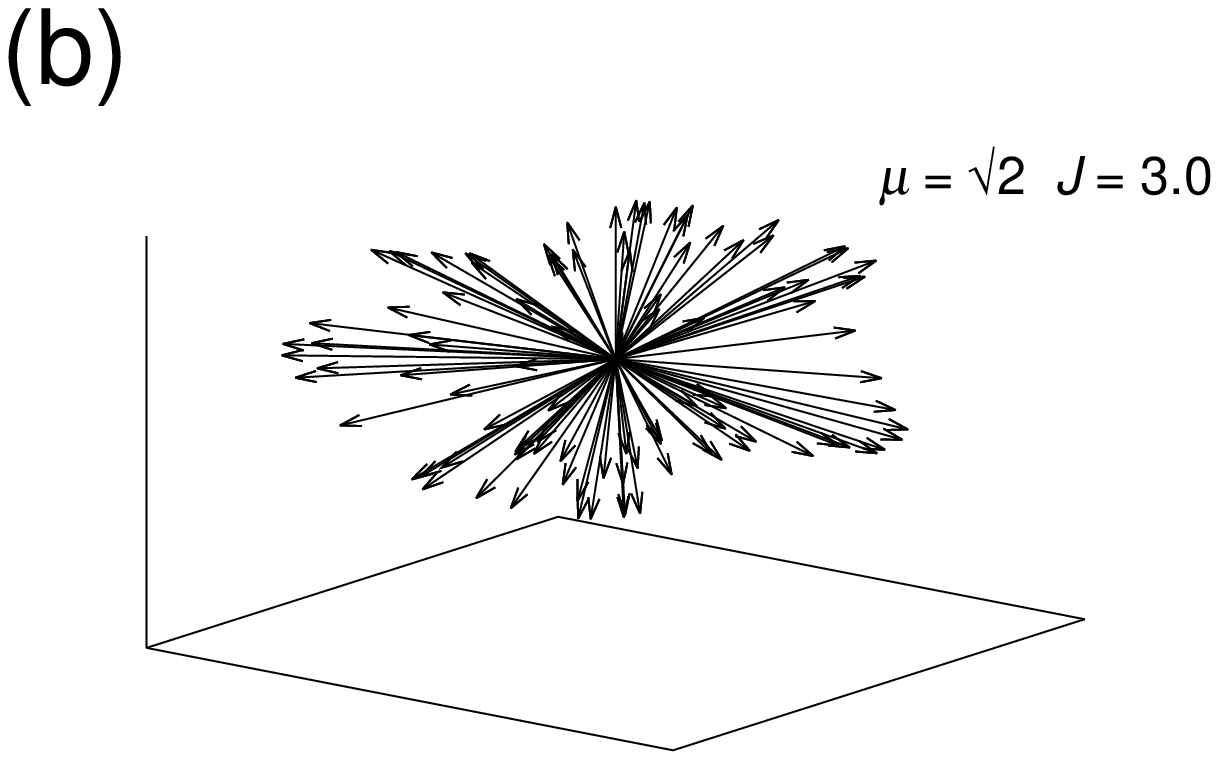}
 \caption{The low-temperature localized-spin configurations of the one-dimensional Kondo-lattice model at $\mu=\sqrt 2$ corresponding to the period-4 or the 3/4-filling, obtained by gradually cooling the system to $T=10^{-5}$. The coupling is (a) $J=0.1$, and (b) $J=3.0$, each lying in the collinear and the noncoplanar chiral regimes, respectively.  The lattice size is $L=100$. In (b), the origin of all the spin vectors is collected to a common point for illustration.
}
\end{figure}
\begin{figure}
 \includegraphics[width=7.5cm]{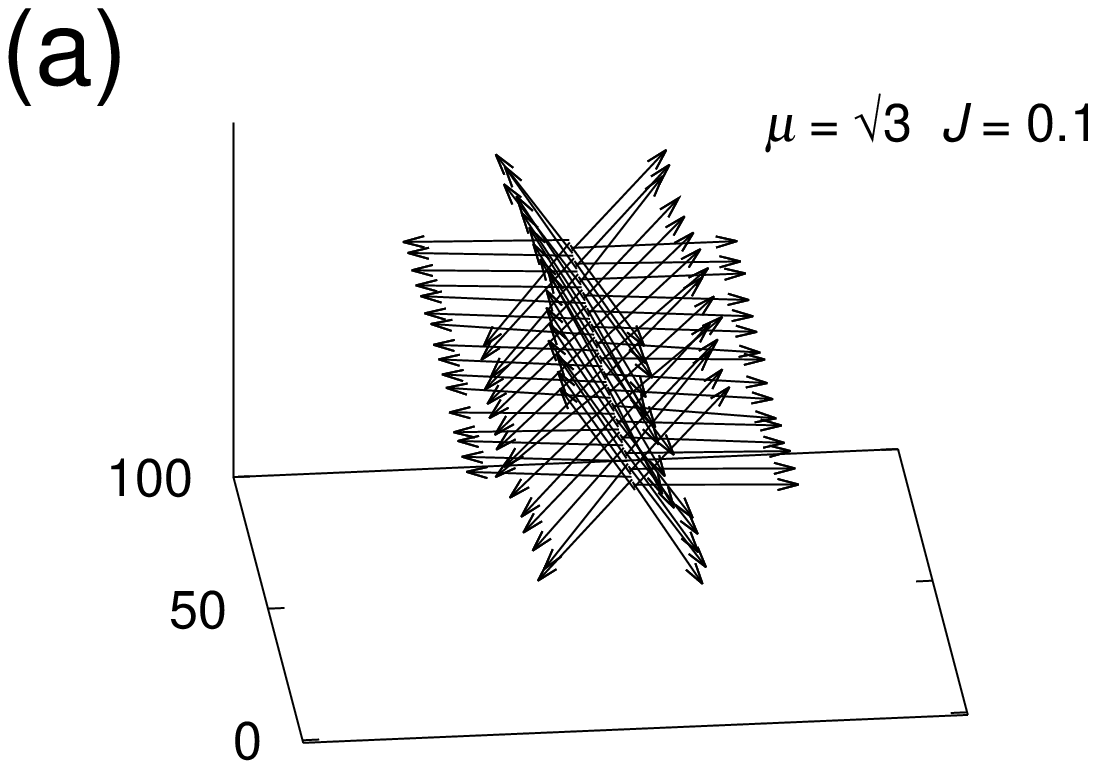}
 \includegraphics[width=7.5cm]{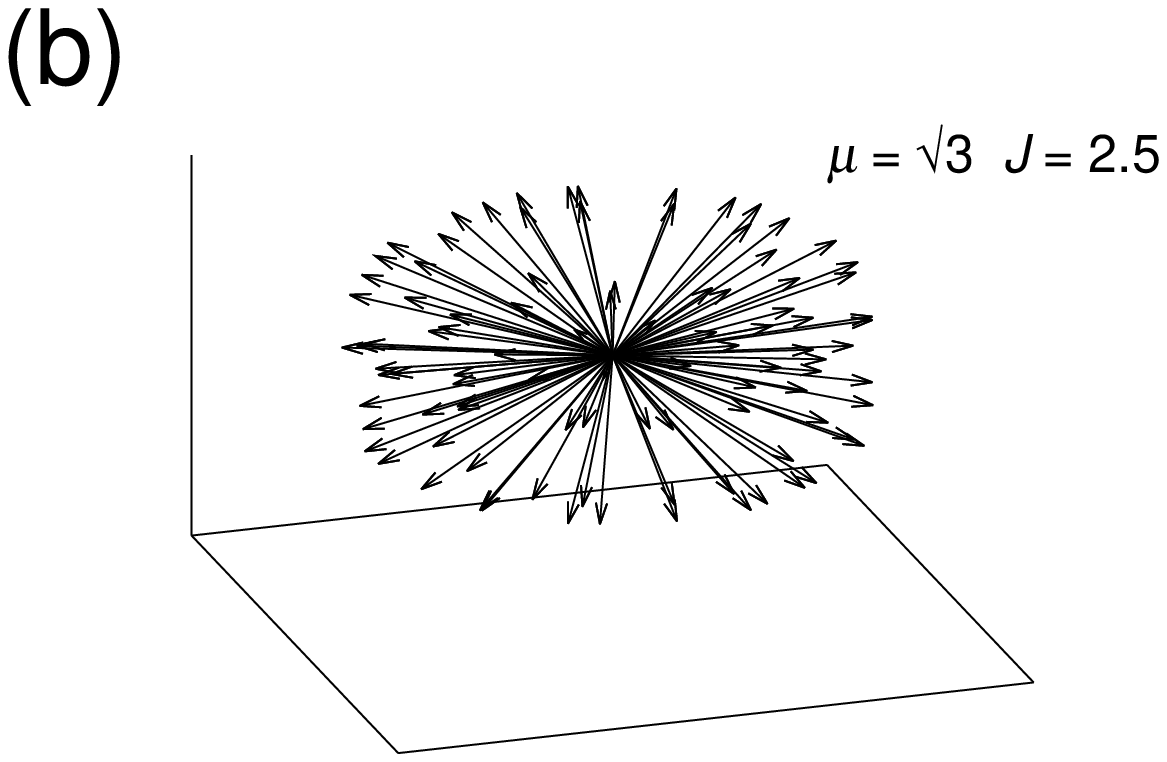}
 \includegraphics[width=7.5cm]{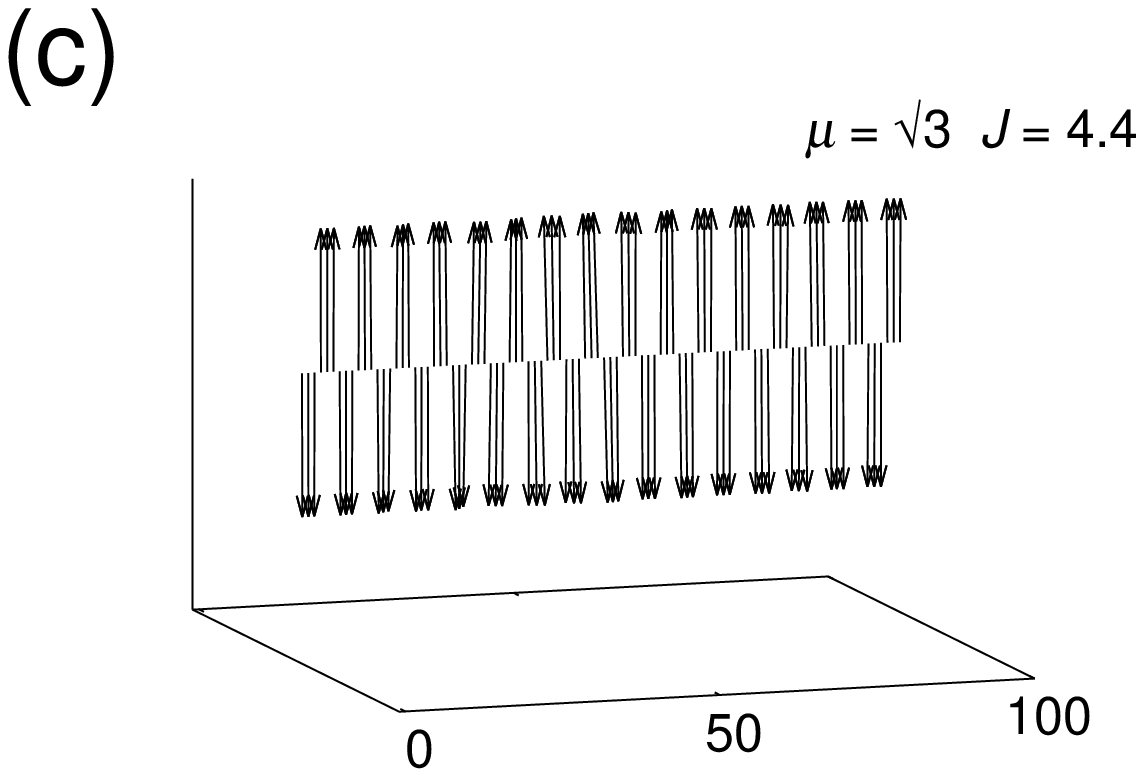}
 \caption{The low-temperature localized-spin configurations of the one-dimensional Kondo-lattice model at $\mu=\sqrt 3$ corresponding to the period-6 or to the 5/6-filling, obtained by gradually cooling the system to $T=10^{-5}$. The coupling is (a) $J=0.1$, (b) $J=2.5$, and (c) $J=4.4$, each lying in the coplanar helical, the noncoplanar chiral, and the collinear regimes, respectively.  The lattice size is $L=96$. In (b), the origin of all the spin vectors is collected to common a point for illustration.
}
\end{figure}

\subsection{The gap opening}

 In this subsection, we further examine the situation around the period-4 region at $\mu=\sqrt 2$ where the collinear ``up-up-down-down'' structure, never realized in the corresponding RKKY model, is stabilized even in the weak-coupling limit of $J\rightarrow 0$, thereby apparently invalidating the perturbative approach. 
 
 For $J=0$, the energy spectrum of the model is given by $\epsilon (k) = -2\cos k$. With the period-4 in mind, we show in Fig.8(a) this dispersion curve in the reduced Brillouin zone associated with the 4-sublattices, [$0,\frac{\pi}{2}$]. The energy band consists of eight branches, but the two associated with the up and the down spins are degenerate for $J=0$. At the 1/4- or the 3/4-filling, a gap might open at $k=\pi/4$. Then, we estimate the effect of a small nonzero exchange coupling $J$ at and around $k=\pi/4$ by diagonalizing the full Hamiltonian in the space of certain period-4 localized-spin patterns, two of which might be the collinear ``up-up-down-down'' state and the coplanar 90-degrees spiral state. In the former collinear case, a gap opens at $k=\pi/4$, whereas in the latter coplanar case, a gap does not open, as demonstrated in Fig.8(b). If one calculates the total energy corresponding to these two spin patterns for, say, $J=0.05$, one gets $-3.022177$ for the collinear ``up-up-down-down'' state versus $-3.022152$ for the coplanar state, revealing that the collinear state has a slightly lower energy (by $\sim 0.08\%$) and is stabilized in the $J\rightarrow 0$ limit. The result is consistent with our numerical observation at $\mu=\sqrt 2$. The fact that an infinitesimal $J$ leads to the gap opening invalidates an implicit assumption underlying the standard perturbation approach. Thus, the gap opening is the cause of the failure of the RKKY-model description of the KLM at the 3/4- and 1/4-fillings even in the weak-coupling limit.

 Similar analysis can also be made for other commensurate periods including the period-6. Indeed, in the case of the period-6, the computed total energy of the collinear ``up-up-up-down-down-down'' state turns out to be higher than that of the coplanar 60-degrees spiral state for smaller $J$, but becomes lower in the region around $J\sim 4$ where the collinear state is stabilized, consistently with our numerical observation. 

 We note that the standard AF order realized at half-filling $n=1$ is nothing but the period-2 collinear state, where the gap opens at $k=\pi/2$. In this case, and in this case only, the collinear state coincides with a coplanar state, the 180-degrees spiral, so that the state is realizable also in the RKKY model. The gap energy at $k=\pi/2$ can easily be estimated to be $\frac{J}{2}$. By matching the chemical potential $\mu$ with this gap energy, one gets the relation $J=2\mu$, which roughly reproduces the numerically determined phase boundary into the AF phase.

 While there is no special reason to expect a gap opening for general incommensurate wavenumbers $k$, the gap might open for commensurate $k$. We have checked that the gap indeed opens for the general period-$n$ collinear spin structure including odd $n$. For any period other than 4, however, the total energy turns out to be higher for the collinear spin structure than for the coplanar spiral structure of a turn angle $\frac{2\pi}{n}$. This means that, in the weak-coupling limit $J\rightarrow 0$, the collinear spin structure is stabilized only in the case of the period-4, consistently with our numerical observation. 

\begin{figure}
 \includegraphics[width=8cm]{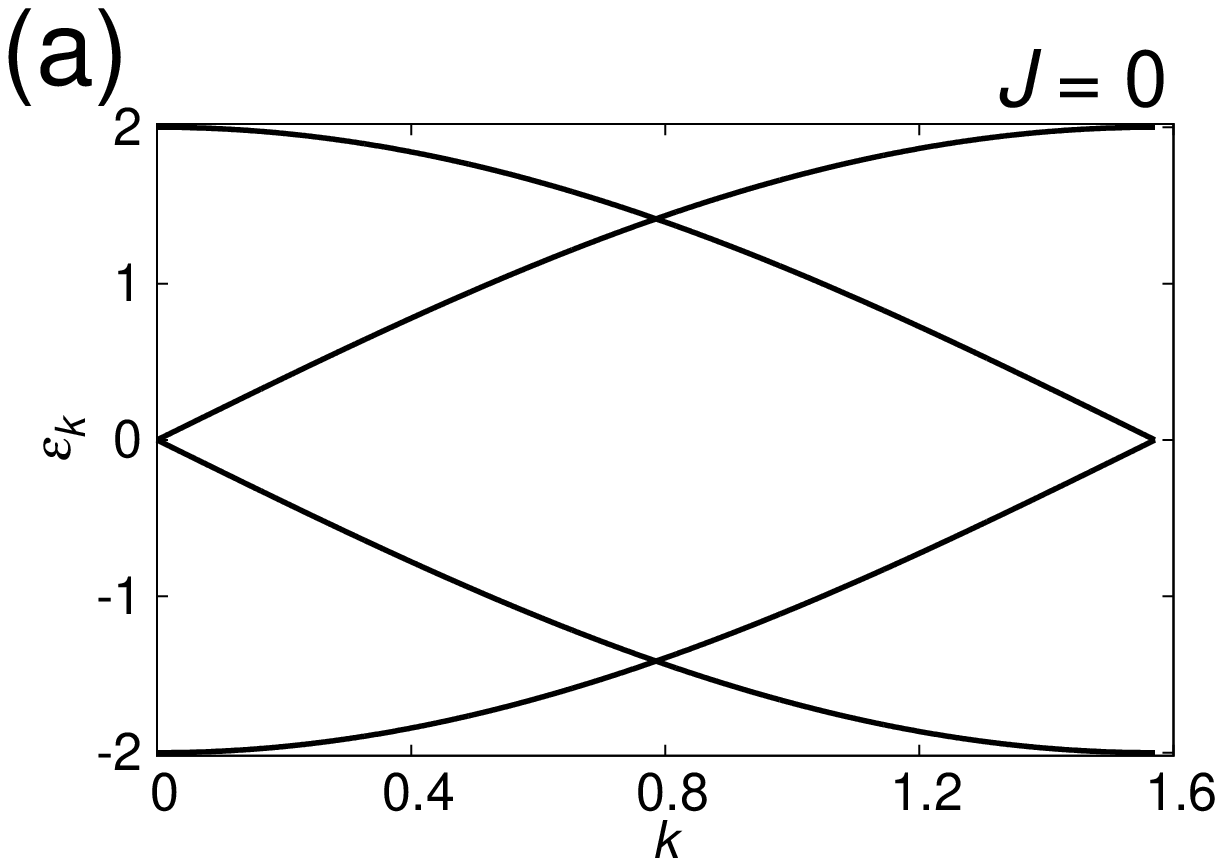}
 \includegraphics[width=8cm]{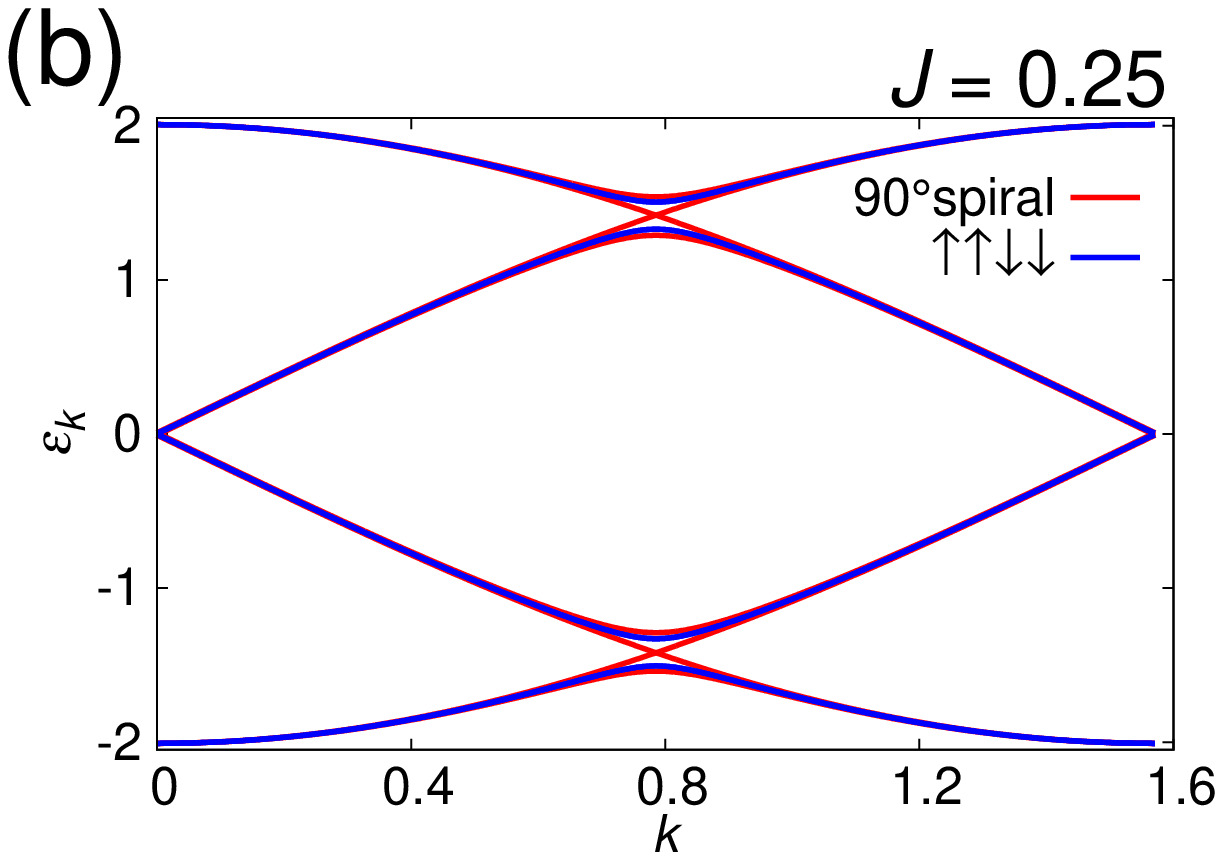}
 \caption{(color online). The band structure of the one-dimensional Kondo-lattice model in the reduced Brillouin zone of the period-4 for the case of (a) $J=0$, and (b) $J=0.25$. In (b), the energy spectrum associated with the collinear ``up-up-down-down'' structure and the one associated with the coplanar 90-degrees spiral structure are shown. The gap opens at $k=\pi/4$ only for the former case.
}
\end{figure}

\subsection{The spin structure factor and the chirality}

 In this subsection, we further investigate the low-temperature spin structures at each phase, especially at the chiral noncoplanar phase, by computing the spin structure factor and the scalar spin chirality. The spin structure factor $F_s(k)$ is defined by $F_s(k)=|\vec S(k)|^2$ where $\vec S(k)$ is a Fourier transform of the spin $\vec S_j$, $\vec S(k)=\frac{1}{\sqrt N} \sum_j \vec S_j e^{ikj}$.

 The $F_s(k)$ computed for the incommensurate coplanar spiral is shown in Fig.9, the corresponding real-space spin structure being given in Fig.5(a). It exhibits a sharp peak at a $k$-value associated with the spiral wavenumber (paired with a peak at $k=2\pi-k$). In the example of Fig.9, the peak appears at $k=0.78\pi\simeq 2.450442$ and $1.22\pi \simeq  3.832743$, close to the expected incommensurate $k$-values of $k\simeq 2.532207$ and $\simeq 3.750978$. Note that, in finite systems under the periodic boundary condition, possible $k$-values are limited as multiples of $2\pi/L$. The peak is sharp, indicating that the realized spin configuration is the standard spiral characterized by a definite wavenumber.

 For the spiral state at $\mu=\sqrt 3$ corresponding to the period-6, $F_s$ exhibits sharp peaks at commensurate wavenumbers $k=\pi/3$ and $5\pi/3$ as expected (the data not shown here). For the spiral states at other $\mu$-values, the situation is qualitatively the same either for a commensurate or an incommensurate spiral.

\begin{figure}
 \includegraphics[width=7.5cm]{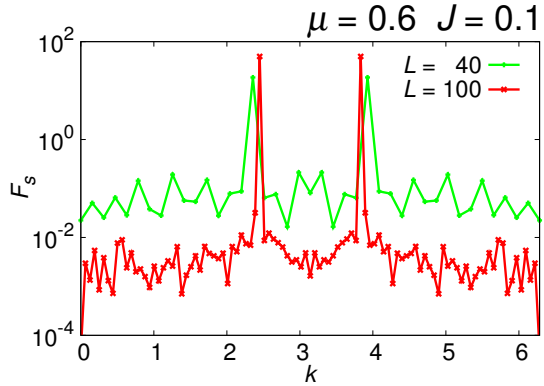}
 \caption{(color online). The spin structure factor of the incommensurate coplanar spiral state at $\mu=0.6$ and $J=0.1$. The lattice size is $L=40$ and 100.
} 
\end{figure}

 For the collinear ``up-up-down-down'' and ``up-up-up-down-down-down'' states, the computed $F_s$ also exhibits a sharp peak at the wavenumber $k=(\pi/2, 3\pi/2)$ and $k=(\pi/3, 5\pi/3)$, respectively, with an additional sub-peak appearing at $k=\pi$ in the latter case.

 The $F_s$ for the chiral noncoplanar states are shown in Fig.10(a)-(c), the corresponding real-space spin structures being given in Figs.5(b), 6(b) and 7(b), respectively. The computed $F_s$ exhibits peaks with its position close to those of the coplanar spiral or of the collinear structure, whereas the peak width is considerably broader. For example, in the case of $\mu=0.6$ shown in Fig.10(b), main peaks appear at the positions close to those of the coplanar spiral of Fig.9, with additional sub-harmonics peaks. Meanwhile, each peak has a considerable amount of width. This finite width of the $F_s$-peak is supposed to give rise to the apparently complex incommensurate real-space spin structure shown in Fig.5(b). In the case of $\mu=\sqrt 2$ shown in Fig.10(c), in particular, the $F_s$-peaks of are located considerably away from the $k$-values expected for the period-4, $\pi/4$ and $3\pi/4$, and each peak has a still broader width. 

\begin{figure}
 \includegraphics[width=7.5cm]{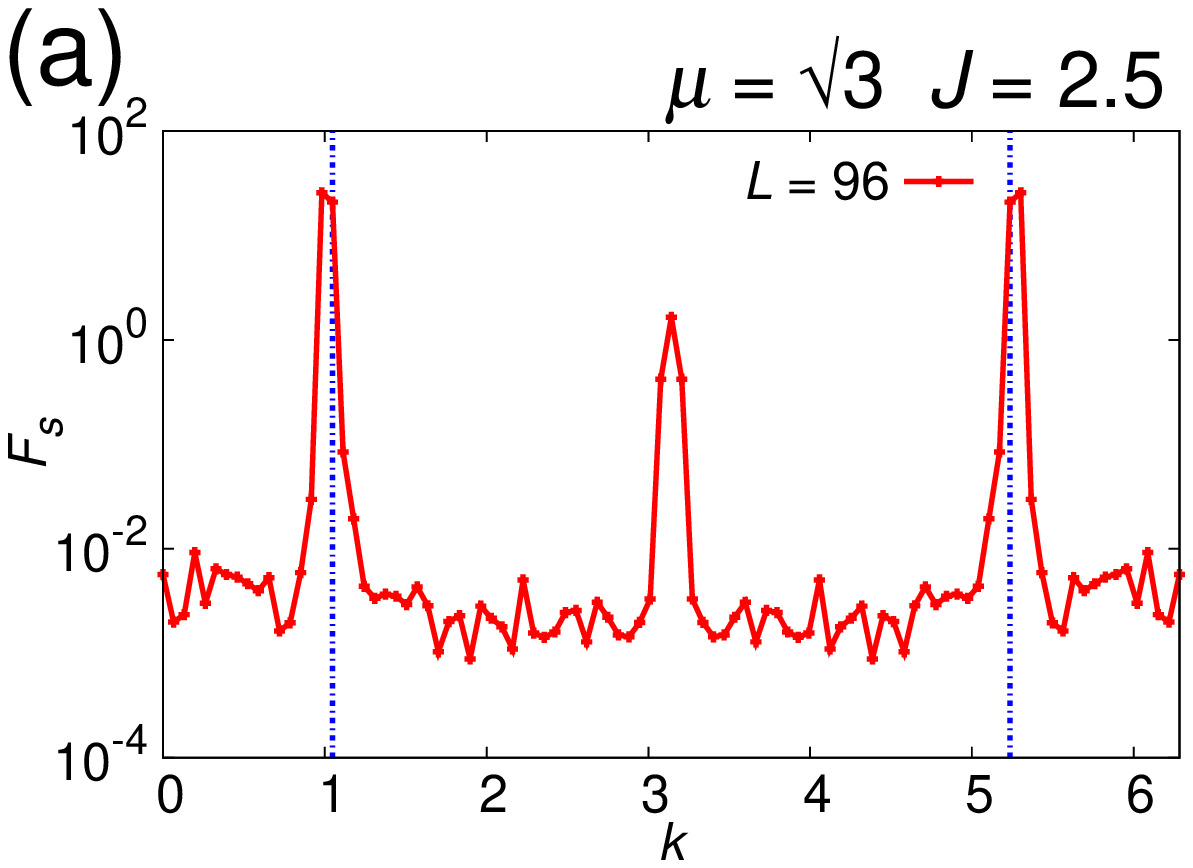}
 \includegraphics[width=7.5cm]{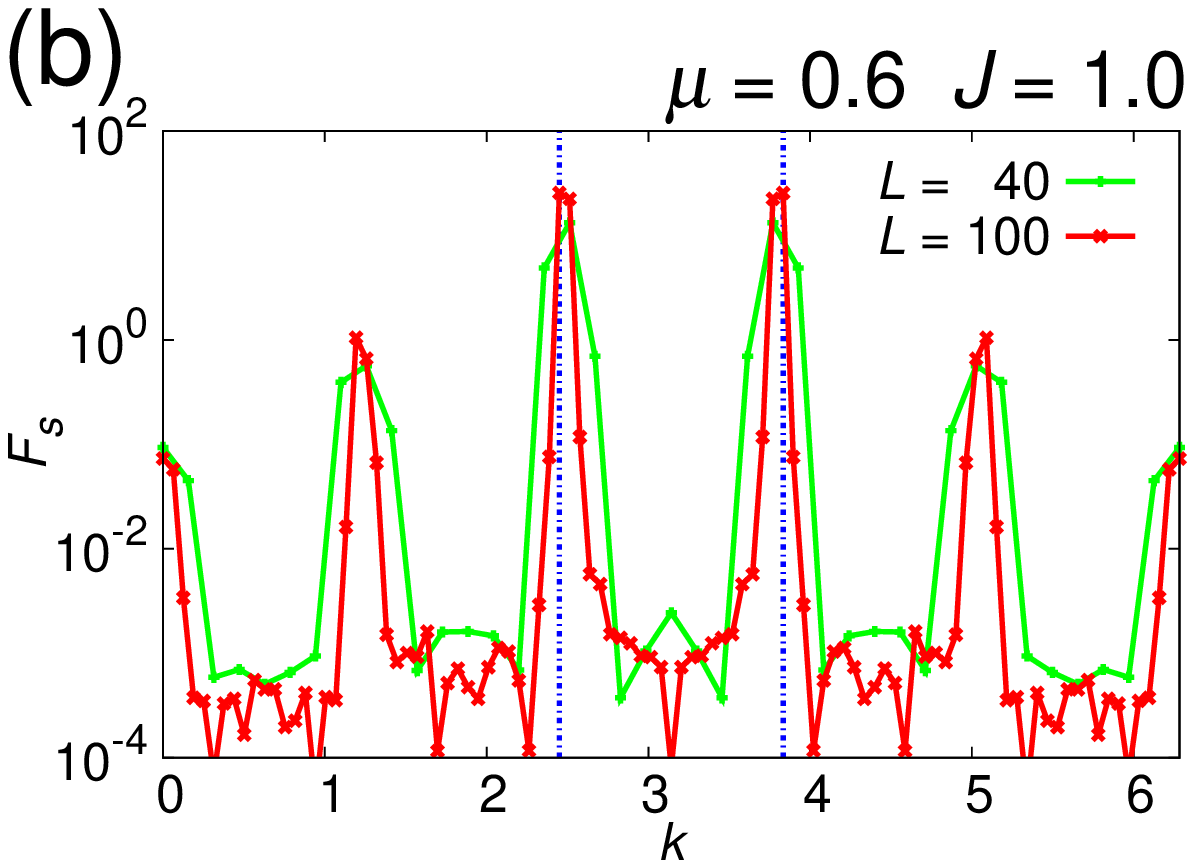}
 \includegraphics[width=7.5cm]{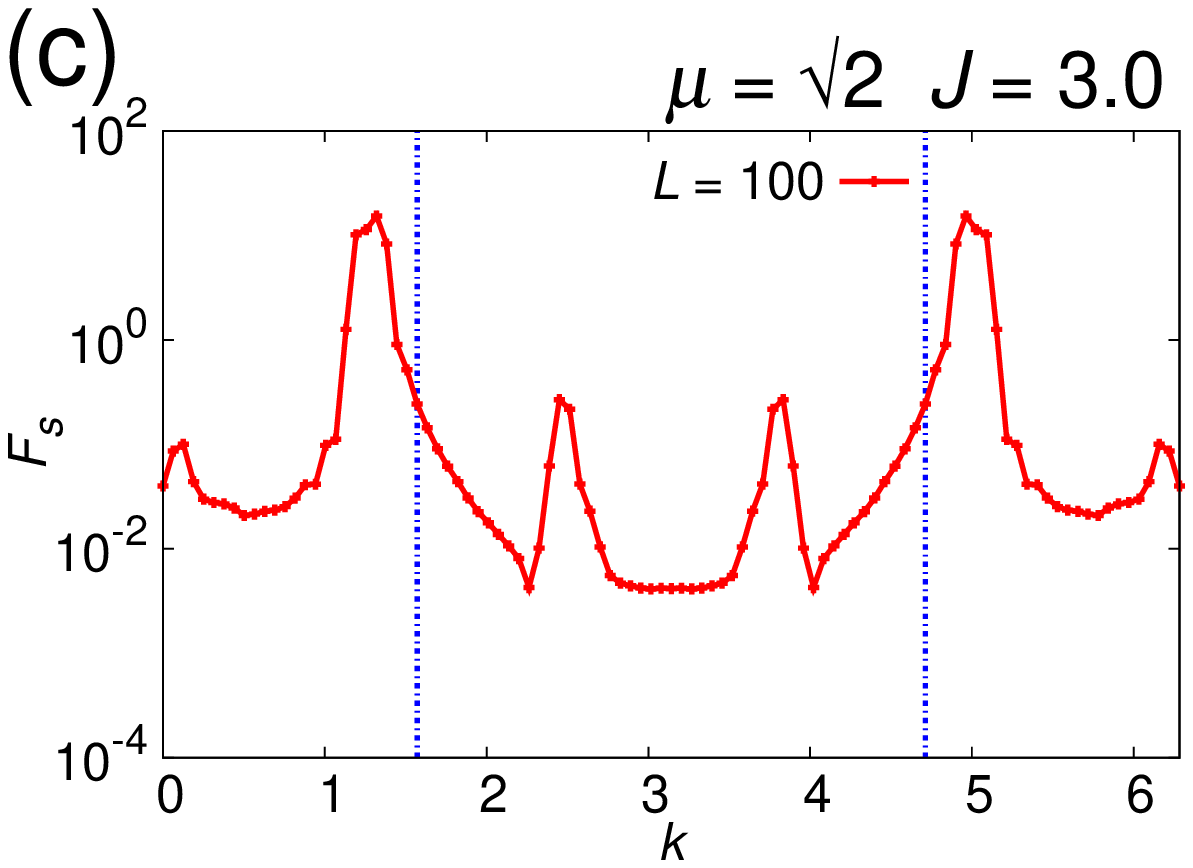}
 \caption{(color online). The spin structure factor of the chiral noncoplanar states at (a) $\mu=\sqrt 3$, $J=2.5$, (b) $\mu=0.6$, $J=1.0$, and (c) $\mu=\sqrt 2$, $J=3.0$. The lattice size is $L=96$ for (a), $L=40,100$ for (b), and $L=100$ for (c). In (a) and (c), the $k$-values corresponding to the expected periodicity, {\it i.e.\/}, the period-4 (a) and the period-6 (c), are indicated by the dashed vertical lines, while, in (b), the $k$-value corresponding to the peak position of the coplanar spiral given in Fig.9 for $J=1.0$ and $L=100$ is indicated by the dashed vertical line.
} 
\end{figure}

 To get further information on the nature of the chiral noncoplanar spin states, we also compute the scalar spin chirality. The scalar spin chirality $\chi_i$ may be defined by the three-spin product,
\begin{equation}
\chi_i =  \vec S_{i-1}\cdot \vec S_i\times \vec S_{i+1} , 
\end{equation}
\begin{figure}
 \includegraphics[width=7.5cm]{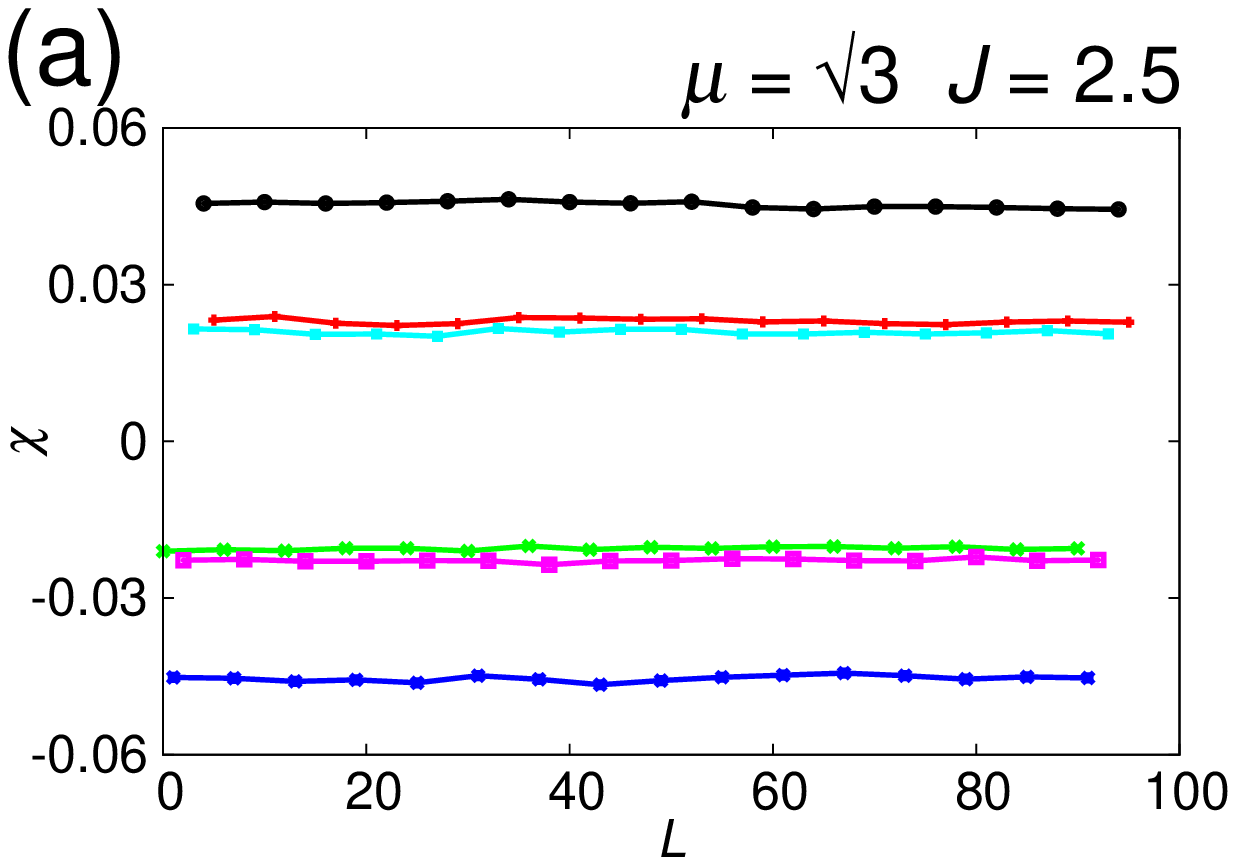}
 \includegraphics[width=7.5cm]{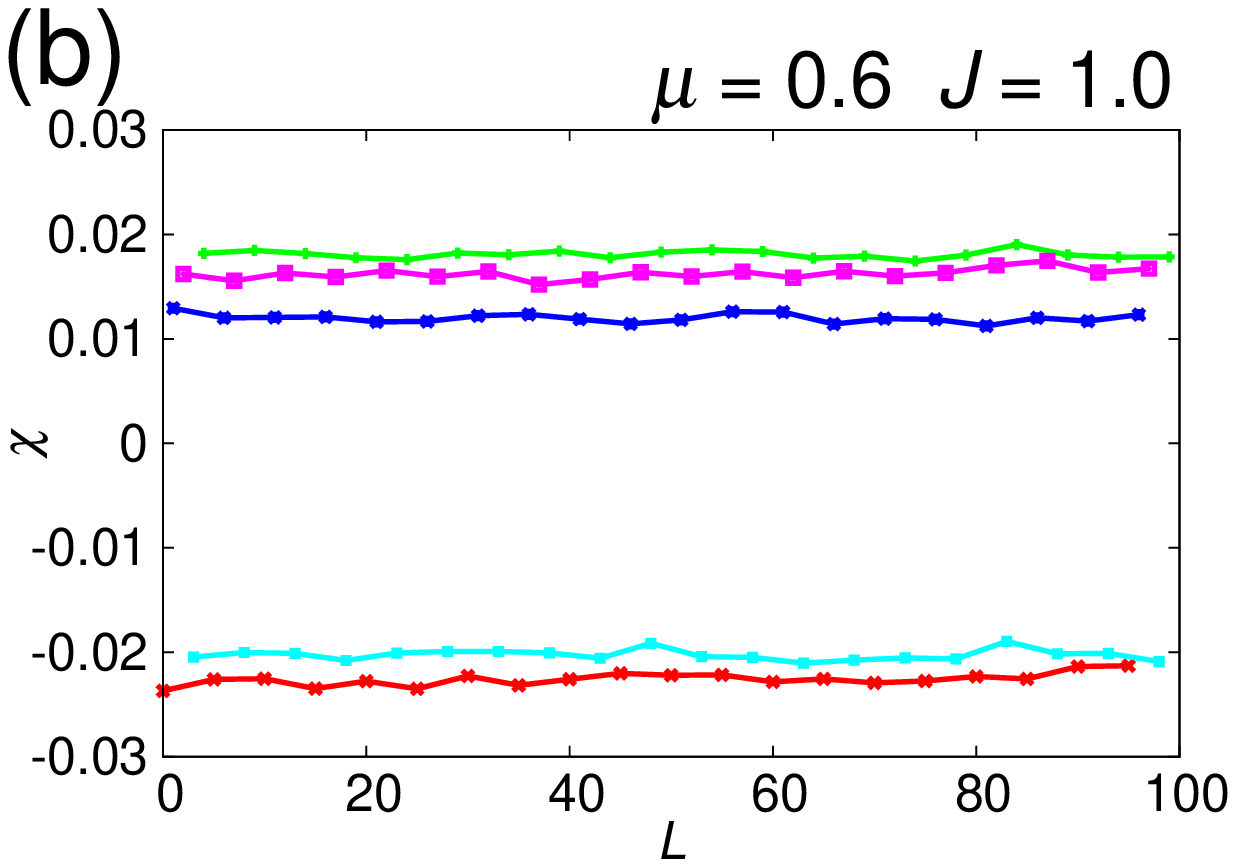}
 \includegraphics[width=7.5cm]{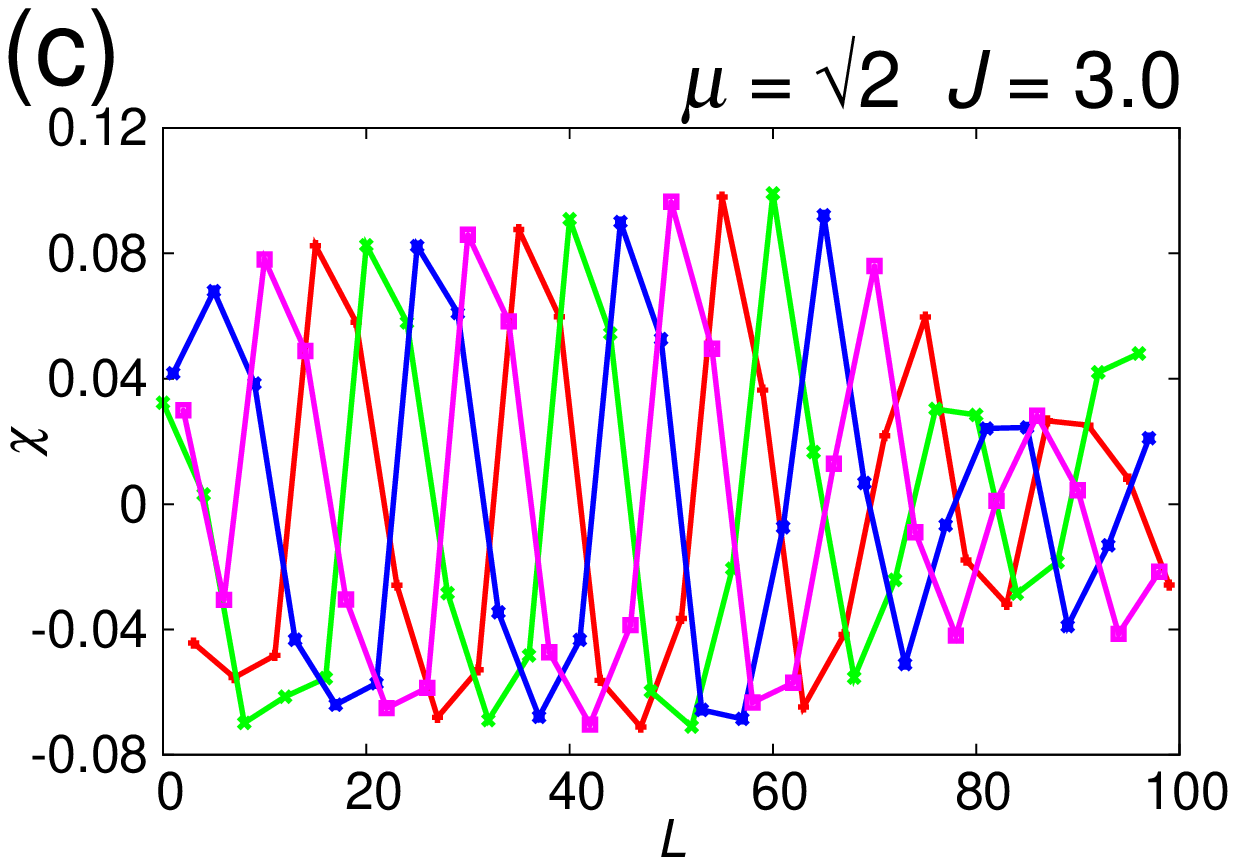}
 \caption{(color online). The position-$i$ dependence of the scalar spin chirality $\chi_i$ for the cases of (a) $\mu=\sqrt 3$, $J=2.5$, (b) $\mu=0.6$, $J=1.0$, and (c) $\mu=\sqrt 2$, $J=3.0$. In (a), the data at every 6 sites are connected by lines to demonstrate the period-6. Likewise, every 5 sites in (b) and every 4 sites in (c) are connected by lines. In (c), the expected period-4 is not observed.  The lattice size is $L=96$ for (a), and $L=100$ for (b) and (c).
} 
\end{figure}

The position-$i$ dependence of the scalar chirality $\chi_i$ at a low-temperature is shown in Fig.11 for the cases of (a) $\mu=\sqrt 3$, (b) $\mu=0.6$, and (c) $\mu=\sqrt 2$, each corresponding to the noncoplanar spin structures shown in Figs.7(b), 5(b) and 6(b), respectively. For $\mu=\sqrt3$, the scalar chirality exhibits a clear periodicity of the period-6 as can be seen from Fig.11(a), even though such a periodicity is not eminent in the real-space spin configuration of Fig.7(b). A similar periodicity is observed also for $\mu=0.6$ shown in Fig.11(b), where the chirality exhibits the period-5 periodicity. Interestingly, for the coplanar spiral realized at $\mu=0.6$, the $F_s$-peak appears not exactly at the period-5 position, but slightly away from it as shown in Fig.9, leading to the incommensurate spiral of Fig.5(a).
%Interestingly, $\mu=0.6$ is rather close to $2\sin \frac{\pi}{10}\simeq 0.6082 \cdots $ corresponding to the 3/5-filing ($n=\frac{6}{5}$).

 The period-4 case at $\mu=\sqrt 2$, shown in Fig.11(c), again turns out to be somewhat special. The expected period-4 tendency is never discernible here, in sharp contrast to the other cases shown in Figs.11(a) and (b). Thus, the period-4 seems always special both in the weak- and intermediate-coupling regimes.

\section{Summary and discussion}

On the basis of the MC method combined with the ED technique, we investigated the low-temperature magnetic properties of the 1D KLM with classical localized spins, with special attention to the  $T\rightarrow 0$ spin structure and the phase diagram, including the weak-, intermediate- and strong-coupling regimes.

In addition to the standard F, AF and coplanar helical phases, we uncovered several new phases such as the collinear phases of the period-4 (``up-up-down-down'' state) and 6 (``up-up-up-down-down-down'' state) and the chiral noncoplanar phase. The chiral noncoplanar phase and the period-6 collinear phase are stabilized in the intermediate coupling regime, while the collinear ``up-up-down-down'' phase is stabilized toward the weak-coupling regime. Especially, just at the 1/4- and 3/4-fillings, it is stabilized even in the weak coupling $J\rightarrow 0$ limit. 

 We also investigated the properties of the 1D RKKY classical Heisenberg model known to be perturbatively derived as an effective spin Hamiltonian in the weak-coupling limit $J\rightarrow 0$. The low-temperature spin state of the RKKY model turned out to be a coplanar spiral of the wavenumber $k_F$. Comparison with the spin configuration of the 1D KLM revealed that the perturbation approach failed at the 1/4- and 3/4-fillings, where the collinear ``up-up-down-down'' state, never stabilized in the RKKY model, was stabilized in the KLM even in its weak-coupling limit $J \rightarrow 0$. The gap opening was found to be the cause of this failure.

 Whether such a failure of the perturbative scheme, and the stabilization of the  collinear state of the non-perturbative character, is specific to 1D, or has some counterparts in higher dimensions \cite{Hayami,Agterberg}, deserves further clarification. In this context, it might be interesting to point out that, for the 3D simple-cubic lattice at the 1/4-filling, the recent calculation suggested that the four-sublattice commensurate noncoplanar state (the ``triple-$Q$ state'') was stabilized \cite{Hayami}.  

 The stabilization of the chiral noncoplanar state over a wide range of the intermediate-coupling regime, never realizable in the corresponding 1D RKKY Heisenberg model, is also interesting. Though the associated spin configuration looks quite complicated in real space, the periodicity becomes visible when one probes the state via the scalar chirality. The region around the period-4 seems special, since such a hidden periodicity is absent. Further studies are desirable to characterize and classify these noncoplanar chiral states.

 Again, whether the stabilization of the chiral noncoplanar states in the intermediate-coupling regime is due to the speciality of the one-dimensionality, or is more generic to higher-dimensional systems \cite{HayamiPRB}, is an interesting open issue. In this context, several types of chiral noncoplanar states were recently reported in the KLM on frustrated lattices, including the 2D triangular \cite{Martin,Akagi,Kato,Akagi12} and the 3D pyrochlore lattices \cite{Chern,Ishizuka}. In particular, recent studies on the kagome lattice reported a variety of incommensurate noncoplanar states in a wide region of the phase diagram \cite{Ghosh,Barros}. In some cases, the nesting effect associated with the conduction-electron Fermi surface might play an important role,\cite{Martin,Chern} but there is also an occasion that the nesting effect is irrelevant \cite{Akagi}. 

 Our present analysis was restricted to the classical localized spin. The properties of the corresponding quantum-mechanical model are also of interest, which might differ considerably from the classical ones. For example, there exist two different types of F states, no AF state exists even at the half-filling, the spin-liquid state with short-range spin correlations exists in a wider parameter range, apparent absence of the phase separation, {\it etc\/} \cite{Tsunetsugu,McCulloch,Garcia,Basylko,Motome10,Peters}. Hence, even in the simple 1D model, to map out the phase diagram with varying the spin quantum number from $S=1/2$ to $S=\infty$ would be an interesting open issue.

  The authors are thankful to Prof. Y. Motome for useful discussion. This study is supported by Grants-in-Aid for Scientific Research No. 25247064.

\end{document}